\documentclass[12pt]{iopart}
\begin{document}

\title[Dynamics in Semi-Classical Loop Quantum Cosmology]{Early Universe 
Dynamics in Semi-Classical Loop Quantum Cosmology} 

\author{James E. Lidsey}

\address{Astronomy Unit, School of Mathematical 
Sciences,   
Queen Mary, University of London, Mile End Road, LONDON, E1 4NS, U.K.}

\begin{abstract}
Within the framework of loop quantum cosmology, there exists a
semi--classical regime where spacetime may be approximated in terms of  
a continuous manifold, but where the standard Friedmann equations of classical 
Einstein gravity receive 
non-perturbative 
quantum corrections. An 
approximate, analytical approach to studying cosmic dynamics
in this regime is developed for both spatially flat and 
positively-curved isotropic universes sourced by a 
self--interacting scalar field. In the former case, a direct 
correspondence between the classical and semi--classical 
field equations can be established together with a scale factor 
duality that directly relates different expanding and contracting 
universes. Some examples of non--singular, 
bouncing cosmologies are presented together with a scaling, 
power-law solution.

\end{abstract}



\maketitle

\section{Introduction}

The old idea \cite{T1934} of 
bouncing and oscillating universes has attracted renewed interest
in recent years. Leading alternatives 
to inflationary cosmology, such as the pre--big bang scenario 
(for a review, see, e.g., Ref. \cite{pbb,lwc}) and 
the cyclic/ekpyrotic model \cite{cyclic1,cyclic2,cyclic3} developed from 
brane dynamics in heterotic M--theory 
\cite{witten,townsend,hw1,hw2}, both
involve a contracting universe that undergoes a bounce into an 
expanding phase. 

However, the physical process that leads to the bounce  
in these scenarios is unclear. Indeed, 
within the context of classical Einstein gravity, a contracting, 
spatially flat Friedmann--Robertson--Walker (FRW) universe 
can only undergo a non--singular bounce into an expansionary 
phase if the null energy condition is violated \cite{null}.
Exotic forms of `phantom' matter with negative kinetic energy
are often invoked to overcome this shortcoming of the classical theory
\cite{gibbons,aw,bkm}, even though this type of matter 
leads to negative norm states at the quantum level. 
Quantum gravitational effects can also 
lead to non--singular
cosmologies. Models inspired by string/M--theory include 
those where dilatonic loop and higher--order curvature corrections to the 
tree--level action are introduced \cite{higher,high1,high2}. 
A number of bouncing braneworld scenarios have also been developed
\cite{bb,bb1,bb2,bb3,bb4}. 

An alternative leading candidate for a non--perturbative 
and background--independent theory of quantum gravity 
is loop quantum gravity (LQG).   
(For reviews, see, e.g., Ref. \cite{loopreview,loop1}).
This is a canonical quantization of Einstein gravity 
in terms of Ashtekar variables \cite{ashtekar}. 
Loop quantum cosmology (LQC) is the restriction of LQG to 
symmetric states. Below a critical value of the scale factor, $a_i$,
the discrete nature of spacetime is important and the Hamiltonian 
constraint is a difference equation for the wavefunction 
of the universe \cite{bojowald4}. 
Classical dynamics is recovered when the size of the 
universe exceeds a second critical value, $a_*$. Depending on the
quantization scheme employed, there is an intermediate `semi--classical' 
regime, $a_i < a \ll a_*$, 
where the dynamics of the universe is determined by coupled, 
ordinary differential equations (ODEs), but where non--perturbative
quantum corrections are important. It was recently shown that  
these corrections generically 
lead to non--singular bouncing and oscillatory behaviour 
\cite{loopbounce,lb1,lb2,lb3,dh}.
 
In view of the above developments, 
the purpose of the present work is to develop an approximate 
analytical framework for studying the dynamics    
of FRW cosmologies within the context of 
semi--classical LQC when the universe is sourced by a 
self--interacting scalar field. In Section 2, we 
briefly review the main features of LQC, highlighting the quantization
ambiguities that lead to important 
phenomenological consequences. Section 3 develops the analytical 
framework by identifying the scalar field as a natural dynamical variable 
of the system. In the case of spatially 
flat models, a formal correspondence
is established between the quantum--corrected Friedmann 
equations and those of classical Einstein gravity. 
This enables a `scale factor duality' that 
relates different backgrounds through an inversion of the 
scale factor to be identified. This duality 
can be extended to spatially-curved cosmologies.
Classes of non--singular bouncing cosmologies are found in Section 4, 
together with a power-law (scaling) solution. We conclude with a 
discussion in Section 5.

\section{Dynamics in Loop Quantum Cosmology}

The physical classical variables in 
isotropic cosmology are the triad component $ |p| =a^2$
and the connection component $c = \frac{1}{2} (k- \gamma \dot{a})$, 
where $\gamma \approx 0.274$ is the Barbero--Immirzi 
parameter \cite{biparameter,mei}
and $k=(0,1)$ for a (topologically compact) 
spatially flat or positively--curved 
universe, respectively\footnote{LQG is a canonical quantization and  
therefore requires a Hamiltonian formulation of the dynamics. 
The negatively--curved FRW cosmology represents the isotropic 
limit of the Bianchi type V universe and, since this is an example of 
a Bianchi class B model \cite{em}, its evolution can not be described in 
terms of a standard Hamiltonian treatment \cite{ash1}. We therefore 
do not consider this model further.}. They form a canonical pair
with Poisson bracket $\{ c, |p| \} = 8\pi \gamma G/3$  
and the dynamics is determined by imposing the Hamiltonian constraint 
on these variables: 
\begin{eqnarray}
\label{classham}
{\cal{H}} = - \frac{3}{2\pi G \gamma^2} \left[ c(c-k) +(1+\gamma^2) 
\frac{k^2}{4} \right] \sqrt{|p|} + {\cal{H}}_{\phi}
 =0 
\\
{\cal{H}}_{\phi} = \frac{1}{2} a^{-3} p^2_{\phi} +a^3 V(\phi )  ,
\label{matterham}
\end{eqnarray}
where the matter Hamiltonian, ${\cal{H}}_{\phi}$, corresponds to that of 
a minimally coupled scalar field, $\phi$, with a self--interaction  
potential, $V(\phi )$, and canonically conjugate  momentum 
$p_{\phi}= a^3 \dot{\phi}$. 
 
The basic variables in LQG are holonomies along curves in 
space associated with the connection and triad. 
For FRW models, the holonomies are given by 
$h_i = \exp ( c \tau_i ) = \cos \left( c/2 \right) + 2 
\tau_i \sin \left( c/2 \right)$, 
where $\tau_i = -i \sigma_i/2 \in {\rm SU}(2)$ and $\sigma_i$ are the 
Pauli matrices. 
Quantization of constraint (\ref{classham}) 
follows by representing the holonomies as operators and promoting the 
Poisson bracket to a commutator. 
A primary consequence of LQC is that operators for inverse powers of the 
spatial volume have eigenvalues that 
are bounded and do not diverge at zero volume \cite{bojowald1}. 
The matter Hamiltonian (\ref{matterham}) contains a divergent factor of 
$a^{-3}$. This is quantized by starting from the classical 
identity $a^{-3} =  [ 3  (8\pi \gamma G l)^{-1} \{ c, |p|^l \} 
]^{3/(2-2l)}$, where $l$ is a constant,  
and replacing the connection with holonomies \cite{bojowald2}: 
\begin{eqnarray}
\label{classiden}
a^{-3} =  \left[ \frac{1}{4\pi \gamma G l} \sum_I
{\rm tr} \left( \tau_I h_I \{ h^{-1}_I , a^{2l} \} 
\right) \right]^{3/(2-2l)} \nonumber \\
=  \left[ \frac{3}{8\pi\gamma G l j (j+1)(2j+1)}
\sum_I {\rm tr}_j \left( \tau_I h_I \{ h^{-1}_I, a^{2l} 
\} \right) \right]^{3/(2-2l)}  ,
\end{eqnarray}
where the second expression follows 
because any irreducible ${\rm SU}(2)$ representation with 
spin $j$ may be chosen when computing the trace \cite{gr,bojowald3}. 
In general, $j$ takes half--integer
values and $j=1/2$ corresponds to the fundamental representation. 
Eq. (\ref{classiden}) is a classical identity, but inverse 
powers of the scale factor are not needed if $0 < l< 1$. 
 
Quantization results in a 
discrete spectrum for the geometrical density operator, 
$\hat{d}_{j,l} = \hat{a^{-3}}$, that 
is well--defined and can be approximated in terms 
of a continuous function of the scale factor such that 
$d_{j,l}(a) \equiv D_l(q)a^{-3}$, where \cite{bojowald2,bojowald3}
\begin{eqnarray}
\label{defD}
 D_l(q) =
\left\{\frac{3}{2l}q^{1-l}\left[(l+2)^{-1}
\left((q+1)^{l+2}-|q-1|^{l+2}\right)\right.\right. \nonumber\\
 - \left.\left.\frac{1}{1 + l}q
\left((q+1)^{l+1}-{\rm
sgn}(q-1) 
|q-1|^{l+1}\right)\right]\right\}^{3/(2-2l)} 
\end{eqnarray}
and $q=a^2/a_*^2$, $a_*^2= a_i^2 j/3$, $a_i= \sqrt{\gamma} 
\ell_{\rm Pl}$ and 
$\ell_{\rm Pl} = \sqrt{G\hbar}$ is the Planck length. 
We refer to Eq. (\ref{defD}) as the `eigenvalue function'. 
The discrete nature of spacetime becomes important below the 
scale $a_i$. In the limit $a_i < a \ll a_*$, 
the eigenvalue function asymptotes to 
\begin{equation}
\label{approxD} 
 D_l (a) \propto
(3/(1+l))^{3/(2-2l)}(a/a_*)^{3(2-l)/(1-l)} 
\end{equation}
whereas classical behaviour, corresponding to $d_{j,l} \propto a^{-3}$, 
is recovered for $a > a_*$. The eigenvalue 
function is peaked around $a\approx a_*$.

Hence, the geometrical density differs significantly from its classical 
limit\footnote{In the spatially flat classical model, 
the Friedmann equations are invariant under a rescaling of the scale factor, 
$a \rightarrow Ca$, for some arbitrary constant $C$, and it 
is conventional to normalize such that $a=1$ at the 
present epoch. The question then arises as to the physical significance of the 
scale $a_*$. In the quantum theory, 
the primary variable is the volume operator and the 
scale factor arises by taking the cube root. In the quantum theory, 
therefore, a different normalization for the scale factor is adopted
and it is the ratio $a_*/a_i$ that determines the range of
the semi--classical regime \cite{abl}.  
The physical scale is not specified until $a_i$ has been evaluated
and this can only be done at the quantum level.} 
below the critical scale $a_*$. Moreover, 
the freedom in writing the same classical expression (\ref{classiden})  
with different values of $( j, l)$ leads to ambiguities at the quantum level 
that have important cosmological consequences 
\cite{bojowaldinflation,loopbounce,lb1,lb2,lb3,ambig,ambig1}. 
For $j > 3$, there exists a semi--classical 
regime $(a_i < a <a_*)$, where spacetime may be viewed as a continuum
but quantum effects in the density remain important. 
The dynamics in this regime is determined by   
replacing the $a^{-3}$ term in the matter 
Hamiltonian (\ref{matterham}) with $d_{j,l}(a)$ \cite{bojowaldinflation}: 
\begin{equation}
\label{semiclassham}
\hat{\cal{H}} = -\frac{3}{8\pi \ell^2_{\rm Pl}} 
\left( \dot{a}^2 +k^2 \right) a
+ \frac{1}{2} 
d_{j,l} p^2_{\phi} +a^3V =0  ,
\end{equation}
where $p_{\phi} = d^{-1}_{j,l} \dot{\phi}$. 

In the following Section we develop an approximate analytical approach 
to FRW cosmologies
within the context of semi--classical LQC. 

\section{Scalar Field Dynamics in Loop Quantum Cosmology}

The Hamiltonian (\ref{semiclassham}) 
determines the dynamics via the constraint equation 
$\hat{\cal{H}} =0$ and the equations of motion 
$\dot{c} = -\{ c,  {\cal{H}} \}$ and $\dot{p}_{\phi} = 
-\{ p_{\phi} , {\cal{H}} \}$. These take the form 
\begin{eqnarray}
\label{Fsemi}
H^2= \frac{8 \pi \ell_{\rm Pl}^2}{3} \left[ \frac{1}{2D_l} \dot{\phi}^2
+ V \right] - \frac{k^2}{a^2} \\
\label{Rsemi}
\frac{\ddot{a}}{a} = - \frac{8\pi \ell^2_{\rm Pl}}{3} 
\left[ \frac{\dot{\phi}^2}{D_l} \left( 1- \frac{1}{4} 
\frac{d \ln D_l}{d \ln a} \right) -V \right]
\\
\label{fieldsemi}
\ddot{\phi} + 3H \left( 1- \frac{1}{3} \frac{d \ln D_l}{d\ln a} \right)
\dot{\phi} + D_l V_{\phi} =0  ,
\end{eqnarray}
respectively, where $H = \dot{a}/{a}$ defines the Hubble parameter 
and a subscript `$\phi$' denotes differentiation with respect 
to the scalar field.

In general, the dynamics in the semi--classical regime 
is complicated due to the non--trivial form of the eigenvalue function 
(\ref{defD}). An important point, however, is that the 
transition in going from the 
semi--classical $(a \ll a_*)$ to classical $(a >a_*)$
domains is extremely rapid: numerical results indicate that once 
the eigenvalue function approaches unity it does so very quickly and 
the asymptotic power--law (\ref{approxD}) for the 
eigenvalue function remains a good approximation 
until $a \approx a_*$ \cite{ambig1}. We therefore consider the 
dynamics in the region where Eq. (\ref{approxD}) is valid. Such an 
approximation improves as progressively higher energy scales are considered.
It then proves convenient to choose units such that $\ell_{\rm Pl}^{-2} 
= 8\pi$ and to define the constant 
$D_* \equiv [3/(1+l)]^{3/(2-2l)} a_*^{3(l-2)/(1-l)}$. 
Eqs. (\ref{Fsemi})--(\ref{fieldsemi}) are then equivalent to 
\begin{eqnarray}
\label{friedmann}
3H^2 = \frac{1}{2 D_*} \frac{\dot{\phi}^2}{a^r} +V - \frac{k^2}{a^2}\\
\label{Hdot}
\dot{H} = \frac{q}{2 D_*} \frac{\dot{\phi}^2}{a^r}  + \frac{k^2}{a^2}  ,
\end{eqnarray}
where $r \equiv 3(2-l)/(1-l) >6$ and 
$q \equiv (r-6)/6$. 

We now proceed to consider spatially flat and positively--curved
cosmologies in turn.

\subsection{Spatially Flat Cosmologies and Scale Factor Duality}

For spatially flat cosmologies, introducing the new set of variables: 
\begin{equation}
\label{redefine}
b \equiv \frac{1}{a^q} , \qquad 
\frac{d}{d\psi} = \frac{D_*^{1/2}}{q} a^{r/2}(\phi ) \frac{d}{d\phi}, 
\qquad W [\psi (\phi) ] = q^2 V (\phi )
\end{equation}
maps Eqs. (\ref{friedmann})  and (\ref{Hdot}) into the {\em classical}
Einstein field equations
sourced by a minimally coupled, self--interacting scalar field, $\psi$: 
\begin{eqnarray}
\label{classfriedmann}
3 \beta^2 = \frac{1}{2} \dot{\psi}^2 + W \\
\label{classHdot}
\dot{\beta} = - \frac{1}{2} \dot{\psi}^2  ,
\end{eqnarray}
where $\beta \equiv \dot{b}/b = -qH$. The parameters $(b, \beta )$ may be 
viewed as the rescaled scale factor and Hubble parameter, 
respectively.

This formal correspondence between classical and 
(semi--classical) loop quantum cosmological scenarios
has a number of interesting consequences. It represents an 
`ultra--violet/infra--red' duality, in the sense that 
the dynamics of expanding or contracting cosmologies in semi--classical 
LQC can be understood directly by employing methods 
that have been developed previously in the classical scenario. 
In particular, since $q >0$, the correspondence relates a classical expanding 
cosmology with a contracting LQC solution, and vice--versa. 

Moreover, if the scalar field is a monotonically varying function 
of proper time, Eqs. (\ref{classfriedmann}) and (\ref{classHdot}) 
can be transformed into the `Hamilton--Jacobi' form 
\cite{HJ,HJa,HJb,HJc,HJd,HJ1,HJ2}: 
\begin{eqnarray}
\label{HJ1}
W (\psi )= 3\beta^2 -2 \beta^2_{\psi} \\
\label{HJ2}
\beta_{\psi} = -\frac{1}{2} \dot{\psi}  ,
\end{eqnarray}
where a subscript denotes differentiation with respect to the 
rescaled field, $\psi$, and all variables are viewed as explicit functions 
of this parameter. It follows from the definition of the 
$\beta$--parameter that 
\begin{equation}
\label{HJ3}
b_{\psi}\beta_{\psi} = - \frac{1}{2} b (\psi) \beta (\psi) 
\end{equation}
and Eq. (\ref{HJ3}) may then be integrated to yield the rescaled 
scale factor in terms of a single quadrature: 
\begin{equation}
\label{HJ4}
b (\psi ) = \exp \left[ -\frac{1}{2} 
\int^{\psi} d \psi \frac{\beta}{\beta_{\psi}} \right]  .
\end{equation}

Expressing the dynamics in this way allows different expanding and 
contracting backgrounds to be directly related through a 
duality transformation \cite{triality,ruth}. To be specific, 
let us consider a solution to Eqs. (\ref{HJ1}) and (\ref{HJ2}) 
that is parameterized by the functions 
$[ \beta (\psi ), b(\psi ), W (\psi ) ]$ and 
choose as an {\em ansatz} for a new solution $\tilde{\beta} (\psi ) 
= b (\psi )$. Substituting this {\em ansatz} into Eq. (\ref{HJ4}),  
and noting that the function $b(\psi )$ satisfies Eq. (\ref{HJ3}),
then implies that 
the new (rescaled) scale factor and potential are given by  
\begin{eqnarray}
\tilde{b} (\psi ) = \beta (\psi ) \\
q^2 \tilde{V} (\psi ) = 3 b^2 (\psi ) -2 b^2_{\psi} (\psi )  ,
\end{eqnarray}
respectively. On the other hand, 
substituting an {\em ansatz} of the form $\hat{\beta}(\psi ) = 
1/ b(\psi )$ into Eq. (\ref{HJ4})  
implies that the dual scale factor is 
given by $\hat{b} (\psi ) = 1/ \beta (\psi )$ (up to 
an irrelevant constant of proportionality). It then follows from the 
definitions (\ref{redefine}) that the two new solutions 
are related by a `scale factor duality' of the form 
\begin{equation}
\label{sfd}
\tilde{a}(\psi )  = 1/ \hat{a}(\psi ) ,
\end{equation}
where the 
corresponding potentials are determined by 
\begin{equation}
\label{diffpots}
\tilde{V} (\psi ) 
= \hat{V}(\psi ) b^4 (\psi )   .
\end{equation}
Since this duality relates the 
scale factor to its inverse, it manifestly pairs an expanding 
cosmology with a contracting universe. 
It should be emphasized, however, that 
the scalar field $\psi$ does not represent the physical field 
in the LQC setting. The dependence of both the scale factor and 
potential on the field $\phi$ is determined 
from the definition (\ref{redefine}) such that 
$\phi = D^{1/2}_* q^{-1} \int d \psi a^{r/2} (\psi )$. 

\subsection{Positively--Curved Cosmologies}

For positively--curved models, 
the curvature term in Eq. (\ref{friedmann}) implies that the redefinitions
(\ref{redefine}) can not be employed to map the field equations 
(\ref{friedmann}) and (\ref{Hdot}) onto the classical Einstein system. 
On the other hand, a similar duality to that discussed above can be 
uncovered. Defining an effective energy density 
\begin{equation}
\label{effden}
\sigma \equiv \frac{q^2}{2D_*} \frac{\dot{\phi}^2}{a^r} + q^2 V
\end{equation}
implies that the scalar field equation (\ref{fieldsemi})
can be written in the form
\begin{equation}
\label{rhodot}
\dot{\sigma} = - 3\beta \dot{\psi}^2    .
\end{equation}
For a monotonically varying field, this implies that 
\begin{equation}
\label{rhoprime}
\sigma_{\psi} =-3\beta \dot{\psi}
\end{equation}
and it follows from the definition of the $\beta$--parameter that 
\begin{equation}
\label{defHprime}
3\beta^2 = - \frac{b_{\psi}}{b} \sigma_{\psi}  .
\end{equation}
Defining a new variable $\chi \equiv b^{-2/q}$ and substituting 
Eqs. (\ref{effden}) and (\ref{defHprime}) into Eq. (\ref{friedmann}) 
then implies that the Friedmann equation can be expressed in the form  
\begin{equation}
\label{rhochi}
\chi_{\psi} \sigma_{\psi} - \frac{2}{q} \chi \sigma = - 2k^2 q   .
\end{equation}

It may be verified that Eq. (\ref{rhochi}) reduces to Eq. (\ref{HJ3}) 
in the absence of spatial curvature. Since Eq. (\ref{rhochi}) is invariant 
under the simultaneous interchange $\chi (\psi )
\leftrightarrow \sigma (\psi) $, we may 
conclude that a given 
solution parametrized by $[\chi (\psi ), \sigma  (\psi) ]$ can be mapped onto 
a dual model such that the new scale factor is given by 
$\tilde{a} (\psi ) = \sigma^{1/2}(\psi )$ and the effective energy 
density is given by $\tilde{\sigma} (\psi ) = a^2(\psi )$.

In the following 
Section we present some exact solutions to Eqs. (\ref{friedmann}) 
and (\ref{Hdot}) for spatially flat cosmologies.
  
\section{Inflating and Bouncing Cosmologies}
 
\subsection{Scaling Solution}

The correspondence (\ref{redefine})--(\ref{classHdot}) between 
classical cosmic dynamics and loop--corrected models implies that a 
solution to the former can serve as a seed for generating a new 
solution in LQC. The question of whether such solutions are stable to linear 
perturbations can also be determined directly from the 
stability of the seed solution. 

For example, 
one solution of particular importance in conventional 
cosmology is the power--law model driven by an exponential potential
\cite{lm}: 
\begin{equation}
\label{classsol}
b= t^{2/\lambda^2} , \qquad \psi = \frac{2}{\lambda} \ln t , \qquad 
W= \frac{2(6-\lambda^2)}{\lambda^4} e^{-\lambda \psi}   ,
\end{equation}
where the kinetic and potential energies of the field scale at the same 
rate. 
Employing Eq. (\ref{redefine}) implies that the corresponding
scaling behaviour in LQC is given by  
\begin{eqnarray} 
\label{lqcsol}
a= t^{-2/q\lambda^2} , \qquad \phi =  \frac{2 \lambda D_*^{1/2}}{r} 
t^{-r/q\lambda^2} \nonumber \\
V= \frac{2(6-\lambda^2 )}{q^2\lambda^4} \left( 
 \frac{r}{2 \lambda D_*^{1/2}} \right)^{2q\lambda^2/r} \left| 
\phi \right|^{2q\lambda^2/r}   .
\end{eqnarray}
Eq. (\ref{lqcsol}) represents 
a scaling solution in the sense that the ratio of the effective kinetic 
and potential energies of the field, $\dot{\phi}^2/(a^r V)$, remains 
constant. It is interesting that when the 
LQC modifications are significant, the potential is given by a simple 
power of the field. 

Linear perturbations about 
the classical solution (\ref{classsol}) decay with eigenvalue 
$m=(\lambda^2 -6)/2$ in an expanding universe \cite{hw}. Thus, 
an expanding (contracting) solution with a positive (negative) 
potential is always
stable for $\lambda^2 <6$ $(\lambda^2  >6)$. Similar conclusions 
may therefore be drawn 
for the LQC solution (\ref{lqcsol}): the stable solutions 
are represented by a contracting cosmology, where $\lambda^2 <6$ and 
the field moves up a positive potential away from $\phi =0$; or an 
expanding model,
corresponding to a time--reversal of Eq. (\ref{lqcsol}), 
where $\lambda^2 >6$, the potential is negative and the field moves towards 
$\phi =0$. This latter solution represents a superinflationary 
cosmology. (Note that although 
the solution  
becomes singular as $t \rightarrow 0^-$, the approximation 
(\ref{approxD}) inevitably breaks down before this point is 
attained).

\subsection{Bouncing Cosmologies} 

Eq. (\ref{Hdot}) implies that $\ddot{a} >0$ and it follows that  
a necessary and sufficient condition for a contracting 
spatially flat universe to 
undergo a non--singular bounce is that the first derivative 
of the Hubble parameter should remain finite at the point where 
the sign of $H$ changes\footnote{We assume implicitly that 
the scale factor is positive--definite.}. 
Moreover, the potential of the field 
must be negative at the bounce and, consequently, the  
scalar field must be a monotonically varying function of 
proper time. The framework outlined in Section 3.1 
is therefore ideally suited to determining the dynamics of spatially flat, 
bouncing cosmologies. It proves convenient to 
express Eqs. (\ref{HJ1})--(\ref{HJ2}) in terms of the 
original variables such that 
\begin{eqnarray}
\label{HJO1}
V = 3H^2 - \frac{2 D_*}{q^2} a^r H^2_{\phi}
\\
\label{HJO2}
H_{\phi} = \frac{q}{2 D_*} \frac{\dot{\phi}}{a^r}  .
\end{eqnarray}
The scale factor is then given by 
\begin{equation}
\label{HJO3}
a^r (\phi)  = C + \frac{qr}{2D_*} \int^{\phi} d\phi \frac{H}{H_{\phi}}   ,
\end{equation}
where $C$ is a constant of integration, and the time dependence of the scalar 
field follows by integrating Eq. (\ref{HJO2}):  
\begin{equation}
\label{HJO4}
t =  \frac{q}{2D_*} \int^{\phi} \frac{d\phi}{H_{\phi} (\phi) a^r(\phi )} .
\end{equation}

Eqs. (\ref{HJO1})--(\ref{HJO4}) imply that the dynamics of semi--classical 
LQC is determined once the Hubble parameter, 
$H(\phi )$, has been specified as a function of the scalar field.
The potential and scale factor follow immediately from 
Eqs. (\ref{HJO1}) and (\ref{HJO3}) and the corresponding 
time--dependences of the scale 
factor and scalar field can then be deduced after integrating 
Eq. (\ref{HJO4}) and inverting the result. 

In view of this, we now proceed to derive two classes of 
bouncing models by specifying the dependence of the Hubble 
parameter on the value of the scalar field. 
We begin by considering a Hubble parameter of 
the form 
\begin{equation}
\label{ansatz1}
H(\phi) = B\phi^n   ,
\end{equation}
where $B$ and $n$ are arbitrary constants. 
Integrating Eq. (\ref{HJO3}) implies that the scale factor is 
given by 
\begin{equation}
\label{scalepower}
a^r (\phi) = C+ \frac{qr}{4nD_*} \phi^2 
\end{equation}
and Eq. (\ref{HJO1}) implies that the potential has the form
\begin{equation}
\label{potentialpower}
V(\phi) = - \frac{2CB^2 D_*n^2}{q^2} \phi^{2n-2} - 
\left( \frac{3B^2}{q} \right) \left[ 
1+\frac{r}{6} (n-1) \right] \phi^{2n}   .
\end{equation}

For the case where the integration constant 
$C=0$, it can be verified that Eqs. (\ref{ansatz1}) and 
(\ref{scalepower}) correspond to the scaling solution 
(\ref{lqcsol}) discussed above. 
Eq. (\ref{HJO4}) is also integrable for 
the case $n=1$ and $C \ne 0$ and the solution 
in this case is given by 
\begin{eqnarray}
a (\phi ) = C^{1/r} {\rm sec}^{2/r} 
\left( \sqrt{ \frac{CB^2D_*r}{q}} t \right)
\nonumber \\
\phi (t) = \left( \frac{4CD_*}{qr} \right)^{1/2} \tan 
\left( \sqrt{ \frac{CB^2D_*r}{q}} t \right) 
\nonumber \\
\label{potbounce1}
V(\phi ) = - \frac{2B^2C D_*}{q^2} - \frac{3B^2}{q} \phi^2   ,
\end{eqnarray}
where it is understood implicitly that $-\pi/2 < t < \pi /2$. 
The potential (\ref{potbounce1}) has a simple quadratic dependence 
on the scalar field, and differs from the scaling potential 
only by a negative constant, $-2B^2CD_*/q^2$. However, the effect of 
this constant shift in value is significant:  the solution now
represents a cosmology that contracts to a minimum
radius at $t=0$, at which point the field reaches the maximum of its 
potential, and then bounces into an expanding phase as the field 
rolls down the other side of the potential.  The integration constant $C$ 
determines the shift in the potential and represents the minimal value 
of the scale factor. 

In principle, models with more complicated potentials 
can be derived in a straightforward manner by specifying 
$n \ne 1$, although it is not always possible 
to integrate and invert Eq. (\ref{HJO4}) in these cases.

Finally, a second class of model can be found that represents a 
universe that bounces into an asymptotic phase of 
exponential (de Sitter) expansion.  
It is given by  
\begin{eqnarray}
\label{ansatz3}
H = \frac{A\phi}{\left[ 8+qr \phi^2 \right]^{1/2}}, \qquad 
V= \frac{A^2}{q^2} \left( \frac{3 q^2 \phi^2 -2}{qr\phi^2 +8} \right) 
\nonumber \\
a (t)  = \frac{1}{D_*^{1/r}} 
\cosh^{4/r} \left( \sqrt{\frac{rA^2}{16q}} t \right) , \qquad 
\phi (t) = \sqrt{\frac{8}{qr}} \sinh \left( 
\sqrt{\frac{rA^2}{16q}} t \right)   ,
\end{eqnarray}
where $A$ is an arbitrary positive constant and 
the constant of integration in Eq. (\ref{HJO3}) 
is chosen to be $C=1/D_*$. 
The potential asymptotes to a positive constant for large $|\phi|$ and 
has a single minimum at $\phi =0$. The universe bounces into an expanding 
phase as the field passes through this minimum.

\section{Conclusion}

In this paper, we have considered expanding and bouncing 
universes within the context of semi--classical loop quantum cosmology, 
corresponding to the regime where the cosmic dynamics is determined 
by quantum--modified Friedmann equations. The analytic approach
developed here compliments previous numerical and qualitative studies 
\cite{loopbounce,lb1,lb2,lb3,ambig,ambig1} 
and applies in the limit where the eigenvalue 
function of the inverse volume operator 
has a power--law dependence on the scale factor. 
 
Such a study is important as it 
yields insight into the generic nature of these models, 
as well as providing a framework for 
classifying the different types of possible behaviour that may arise. 
In particular, the modified 
evolution equations for spatially flat models
can be recast, after appropriate 
field redefinitions, into the classical Einstein 
equations for a minimally coupled, self--interacting scalar 
field. A similar correspondence between standard relativistic 
cosmology and the high--energy limit 
of the Randall--Sundrum type II braneworld \cite{RSII}
has also been identified recently \cite{greg,greg1,greg2,cllm} and 
it would be interesting to explore possible relationships 
between these apparently unrelated scenarios further.

Within the LQC framework,
such a correspondence is intriguing because it indicates that 
the dynamics at ultra--high energy scales, where the universe 
is just emerging from a truly discrete quantum phase described 
by a difference equation, can be 
analyzed in terms of low--energy classical models, and vice--versa. 
By expressing physical parameters such as the scale factor 
and Hubble parameter as functions of the scalar 
field, we identified 
a form--invariance transformation \cite{triality,ruth,calcagni} that inverts 
the scale factor of the universe. 
This scale factor duality between two different 
backgrounds directly relates expanding and contracting models
and can be extended to the spatially--curved models. 
 
The scale factor duality considered above is 
different to that arising in quantum 
cosmological models derived from lowest--order 
string effective actions \cite{sqc1,sqc2,sqc3,sqc4,sqc5}. 
In this latter class of models, the duality may be viewed as a discrete  
subgroup of a global ${\rm O}(d,d)$ symmetry of the effective action
that arises after toroidal compactification over $d$ spatial dimensions
if the dilaton potential has a suitable functional form \cite{mv}. Thus, 
the duality of string quantum cosmology
relates different backgrounds arising from the 
same fundamental Lagrangian. In contrast, however, 
the scale factor duality of LQC relates solutions 
arising from inequivalent actions, i.e., the transformation 
(\ref{sfd})--(\ref{diffpots}) acts between 
scalar fields with different self--interaction potentials.
Moreover, the scale factor duality of 
LQC becomes apparent when the field equations are expressed as
functions of an effective scalar field defined by the 
condition (\ref{redefine}), whereas the ${\rm O}(d,d)$ 
symmetry of string cosmology is only apparent
when the degrees of freedom are viewed explicitly as functions of cosmic 
time \cite{mv}. 
A further difference between the 
two approaches is that the field is minimally coupled to Einstein gravity
in LQC, whereas the dilaton is non--minimally coupled to gravity.
In view of the (classical) 
conformal equivalence between scalar--tensor and Einstein 
theories of gravity, therefore, 
it would be interesting to investigate the LQC approach 
within the context of scalar--tensor gravity models such as the 
Brans--Dicke theory. Such an approach could yield further 
insight into the similarities and differences 
between the LQC and string quantum cosmology frameworks.  

Finally, we presented simple models that 
exhibit scaling and bouncing behaviour in the semi--classical regime. 
One of the principle motivations for 
developing non--singular bouncing cosmologies is that 
they provide an alternative to the inflationary scenario  
for the formation of large--scale structure in the 
universe. In principle, density perturbations on scales larger than 
the Hubble radius at the epoch of decoupling can be generated 
during a phase of decelerated contraction \cite{gkst,gratton,aw,scalebran}. 
Since a number of issues regarding the propagation 
of perturbations through the bounce presently remain unresolved 
\cite{unresolve,uns1,uns2,uns3}, it is important to develop non--singular 
bouncing models as these can provide a solvable framework 
for analyzing the evolution of perturbations. Moreover, 
the scaling (power--law) solution  has played a central
role in the standard inflationary scenario and represents one of the
few known cosmologies where the density perturbation spectrum can be 
calculated without invoking the slow--roll approximation 
\cite{lm,exactpert}. It would be interesting to investigate whether 
the scaling solution derived above plays a similar 
role in determining the evolution of density perturbations generated 
in the semi--classical LQC regime. Ultimately, this could lead to  
important phenomenological constraints on LQC inflationary models.

\ack We thank P. Singh for helpful discussions.

\end{document}